\documentclass[twocolumn,showpacs,preprintnumbers,amsmath,amssymb]{revtex4}
      \newcommand{\beq}{\begin{equation}}
      \newcommand{\eeq}{\end{equation}}
      \newcommand{\beqa}{\begin{eqnarray}}
      \newcommand{\eeqa}{\end{eqnarray}}
      
      \newcommand{\nn}{\nonumber}

      \newcommand{\bra}{\left\langle}
      \newcommand{\ket}{\right\rangle}
      
      \newcommand{\del}{\partial}

      \newcommand{\al}{\alpha}
      \newcommand{\be}{\beta}
      \newcommand{\ga}{\gamma}
      \newcommand{\de}{\delta}
      \newcommand{\ep}{\epsilon}
      \newcommand{\ka}{\kappa}
      
      \newcommand{\De}{\Delta}

    \newcommand{\kt}{{\cal K}}
    
    \newcommand{\kh}{{\cal K}_>}
    \renewcommand{\(}{\left(}
    \renewcommand{\)}{\right)}
\newcommand{\coarse}[1]{\left\langle #1 \right\rangle_{>}}

\newcommand{\rnd}[1]{\left[ #1 \right]_{P[\rho]}}

\newcommand{\rndp}[1]{\left[ #1 \right]_{P'[\rho]}}

\usepackage{graphicx}
\usepackage{dcolumn}
\usepackage{bm}
\begin{document}
\preprint{}
\title
{
  Renormalization group for the probability distribution of magnetic 
  impurities in a random-field $\phi^4$ model 
  }
 
 \author{Hisamitsu Mukaida}
 \email{mukaida@saitama-med.ac.jp}
 \affiliation{Department of Physics, Saitama Medical College, 
 981 Kawakado, Iruma-gun, Saitama, 350-0496, Japan
 }
 
\author{Yoshinori Sakamoto}
\email{yossi@phys.ge.cst.nihon-u.ac.jp}
 \affiliation{Laboratory of Physics, College of Science and Technology,    Nihon university,
   7-24-1, Narashino-dai, Funabashi-city, Chiba, 274-8501, Japan
 }

\date{February 7, 2003}

\begin{abstract}
  Extending the usual Ginzburg-Landau theory for the random-field Ising model, 
  the possibility of  dimensional reduction is reconsidered.  
  A renormalization group 
  for the probability distribution of magnetic impurities is applied. 
   New parameters corresponding to the extra $\phi^4$ coupling constants  
   in the replica Hamiltonian are introduced.  Although they do not affect 
   the critical phenomena near the upper critical dimension,  they can 
   when  dimensions are lowered. 
 \end{abstract}


\maketitle

\section{Introduction}
\label{introduction}
The random-field Ising model (RFIM) is the Ising model coupled to a random magnetic field\cite{im}.   
Although it has been extensively studied for about three decades,  there remain many unresolved 
problems\cite{n}, one of which concerns   dimensional reduction\cite{aim}.  

Dimensional reduction claims that the critical behavior of the $d$ dimensional 
RFIM is equivalent to the $d-2$ dimensional 
{\em pure} Ising model.  Following this argument,  phase transition 
does not occur in the three-dimensional RFIM.  However, 
it is rigorously proved that the phase transition does occur in this model, 
so that  dimensional reduction does not hold at least $d=3$\cite{i}.  
Various numerical computations are performed to obtain critical exponents 
   in three dimensions\cite{r,nb,hy,mf}. 
On the other hand, it  is not understood whether dimensional reduction works 
in  other dimensions lower than the upper critical dimension, which is 
believed to be  six.   

Since  standard perturbation for the Ginzburg-Landau (GL) theory of the RFIM 
simply leads to the result consistent with  dimensional reduction,   
other approaches were applied.   
Schwartz {\it et al} proposed  modification of   dimensional reduction, 
which indicates correspondence between the $d$ dimensional RFIM and 
the pure Ising model in $d'=d-2-\eta(d')$ \cite{s}.
 Mezard and Young suggested the replica symmetry breaking  
 of the RFIM by  extrapolation from $1/m$ expansion, 
where $m$ is  the components of the spin\cite{my}.   
Lancaster {\it et al} computed exponents,  paying attention to 
many solutions of mean field equations\cite{lmp}.  
Although these works support the breakdown of dimensional 
reduction in  other  dimensions, it has not yet been settled.

The breakdown of  standard perturbation may be caused by 
overlooked relevant operators.  In $4 + \ep$ dimensions, 
Fisher\cite{f}  and Feldman\cite{fe} pointed out that 
there are infinitely many relevant operators 
in $n$-component random spin models.
Near the upper critical dimension, Br\'ezin and de Dominicis investigated 
 that the GL  Hamiltonian corresponding to the RFIM  has not only the usual $\phi^4$ interaction
\footnote{
A similar Hamiltonian is also employed in Ref.\cite{hks} for the high-temperature expansion of the RFIM 
}
\beq
\sum_\al \phi_\al^4, 
\label{phi4-1}
\eeq
where $\al$ denotes the replica index, 
 but also the following extra  $\phi^4$ interactions\cite{bd}:
\beq
  \sum_{\al,\be}\phi_\al^3 \phi_\be, \ \sum_{\al,\be}\phi_\al^2 \phi_\be^2, 
  \ \sum_{\al,\be,\ga}\phi_\al^2 \phi_\be \phi_\ga, \ 
 \sum_{\al,\be,\ga,\de}\phi_\al \phi_\be \phi_\ga \phi_\de. 
\label{phi4-2}
\eeq  
It means that the space of the coupling constants extends to the five dimensions and 
that  a renormalization-group (RG) trajectory should be considered in  five-dimensional space. 
They claimed that  dimensional reduction does not work even near the upper critical dimension
since the non-trivial fixed point  of ${\rm O}(\epsilon)$ becomes unstable 
in $d=6-\epsilon$. 

However, their analysis has ambiguity that originates from the zero-replica limit. 
In fact, they proposed two different 
limiting procedures and derived two sets of beta functions quantitatively different each other. 

In this paper, we circumvent the ambiguity and reconsider RG flow 
in the extended coupling-constant space. 
To this end, we use RG for the random probability 
distribution that controls random potentials 
including the magnetic field.  
This method  was initiated by Harris and Lubensky\cite{hl,l,m} in order to investigate 
a randomly diluted magnetic system 
with non-magnetic impurities.  In this case, it is confirmed that the replica method
 is consistent with the Harris-Lubensky method\cite{l,gl}.  
Here we extend the random probability distribution adopted in the previous
 literatures\cite{im,aim,g}. It corresponds to taking into account the replica interactions 
 (\ref{phi4-2}) as well as (\ref{phi4-1}).

According to Refs.\cite{hl,l},   the coupling constants in the Hamiltonian are 
extended to 
inhomogeneous random potentials 
that are correlated to each other following some random probability distribution $P$.  
After performing  standard RG in the inhomogeneous potentials, 
the change of  the Hamiltonian can be pushed 
into transformation of  $P$.  
Distribution $P$ is characterized by non-trivial cumulants with some parameters.  
Therefore the change of $P$ defines RG flow in that parameter space. 
We can investigate critical behavior from  RG flow.  An advantage of this method 
compared to the replica method is that any ambiguities originating from limiting procedures 
do not arise. 

This paper is organized as follows: 
we show in the next section  the RG for  probability distribution of the 
Gaussian random magnetic impurities.  We can naturally introduce
 five parameters that completely correspond
to the five coupling constants for the interactions 
(\ref{phi4-1}) and (\ref{phi4-2}) in the replica 
Hamiltonian.  
In Sec. \ref{pert}, we analyze the RG by perturbation and  derive the recursion equations 
of the parameters that specify  random-field distribution.  
We also compare the result with that of the replica method\cite{bd}. 
In Sec.  \ref{redef},  we introduce expansion parameters of physical quantities 
that are convenient for analysis near the upper critical dimension. 
In the language of the replica method, it means to determine the scaling dimensions of 
the operators in (\ref{phi4-1}) and (\ref{phi4-2}).  We show that the interaction (\ref{phi4-1}) 
is the unique relevant operator near the upper critical dimension in 
 all the $\phi^4$ interactions. Then  arguments of  dimensional reduction can survive 
near the upper critical dimension.  However, as the dimensions are lowered,   other 
interactions can become relevant, which  cause the breakdown of the dimensional 
reduction.  This picture is consistent with  high-temperature expansion studied by 
Houghton {\it et al}\cite{hks}.  The last section is devoted to summary and discussion.

\section{Renormalization-group transformation for the random probability distribution}
\label{rgt}
\begin{figure}
\begin{center}
\setlength{\unitlength}{1mm}
\begin{picture}(60, 55)(0,0)
     \put(0,-5){ 
\includegraphics[width=60mm]{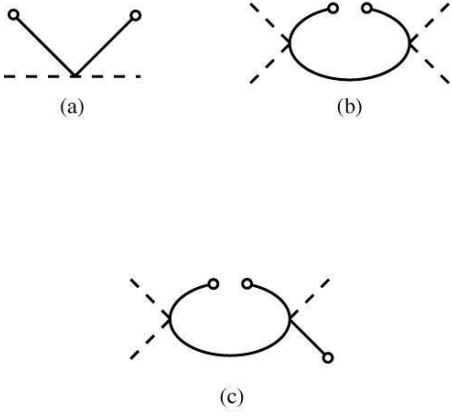}
		}
\end{picture}
\end{center}
\caption{Graphical representations for Eqs. (\ref{de2}), (\ref{de4}) and (\ref{de3}).  
A solid line stands for $G(q)$ and a dashed line for $\phi(p_i)$.
The higher-momentum components of the random field $ {v_1}$ is depicted by an open circle.}
\label{fig_intro}
\end{figure}

In this section,  we formulate the renormalization-group transformation (RGT) for the random probability
 distribution in the presence of magnetic impurities following Refs.\cite{hl, l}.
The GL Hamiltonian  to the RFIM is usually  described by \cite{aim}:
  \beqa
       &&\be H =
       \frac{1}{2} \int_{k_1} \(k_1^2+t\) \phi(k_1)\phi(-k_1) +\int_{k_1} v_1 (k_1) \phi(k_1)
 \nn\\
		&&+ \frac{u_1 }{4!}\int_{k_1, k_2, k_3} \phi(k_1)\phi(k_2) \phi(k_3) \phi(-k_1-k_2-k_3)
\nn\\
 \label{H}
  \eeqa
  where the momenta  $k_1, k_2, k_3$ belong to 
  \beq
   \kt = \left\{ k \ : \ 0 \leq | k | \leq \Lambda \right\}
  \eeq
  and the integral means 
  \beq
    \int_{k_1, .., k_j} \equiv \int_{k_1 \in \kt} \frac{d^d k_1}{(2\pi)^d} \cdots 
	\int_{k_j \in \kt} \frac{d^d k_j}{(2\pi)^d}.  
  \eeq
The random magnetic field $v_1(k)$ follows the Gaussian distribution $P_0[v_1]$ 
  \beqa
    &&P_0 [ {v_1}] \equiv N e^{-\frac{1}{2 \Delta} \int_k  {v_1}(k)  {v_1}(-k) } \nn\\
	&&N \equiv  \(\int {\cal D}  {v_1} e^{-\frac{1}{2 \Delta} \int_k  {v_1}(k)  {v_1}(-k) } \)^{-1}\nn\\ 
	&&{\cal D}  {v_1} \equiv \prod_{k \in \kt} d  {v_1}(k). 
  \label{p0}
  \eeqa
We treat the random field $v_1$ as an external field for a while and examine the RGT.
Namely, we first integrate higher-momentum components of $\phi$ and then perform 
the rescaling of  
the potentials and the fields appropriately.
Let us observe correction terms that appear in  higher-momentum integration.  
Let
$G(q)$ be the free propagator: $G(q) = (q^2 + t)^{-1}$.
By integrating the higher-momentum components by perturbation in $u_1$, we have the following  correction
to the $\phi^2$ term in the leading order:
\beq
   u_1 \int^>_q {v_1}(q)G(q) {v_1}(p_1 + p_2 + q)G(p_1 + p_2 + q) \phi(p_1) \phi(p_2).
\label{de2}
\eeq
Here integration is carried out on higher-momentum space $\kh$ and $\phi(p_i)$ is a 
lower-momentum component.
Similarly, the $\phi^4$ term gets 
\beqa
   &&u_1^2 \int^>_{q_1, q_2}  \de \(\sum_{i=1}^4 p_i + \sum_{j=1}^2 q_j \) 
   \prod_{i=1}^2 {v_1}(q_i) G(q_i) \times
 \nn\\
  &&G(p_1 + p_2 +q_1)\prod_{k=1}^4 \phi(p_i).                                                                                                                                                                                                                                                                                                                                                                                                                                                                                                                                                                                                  \label{de4}
\eeqa
Further,  a cubic interaction emerges through, for example, 
\beqa
  &&u_1^2 \int^>_{q_1, q_2}  \de \(\sum_{i=1}^3 p_i + \sum_{j=1}^3 q_j \) 
   \prod_{i=1}^3 v_1(q_i)G(q_i) \times
 \nn\\
 &&G(p_1 + p_2 + q_1) \prod_{k=1}^3 \phi(p_i). 
 \label{de3}
\eeqa
These corrections are depicted in FIG. \ref{fig_intro}.  Eqs. (\ref{de2}), (\ref{de4}) and (\ref{de3}) 
respectively correspond to (a), (b) and (c) of the figure.  
%
%
The correction terms indicate that the coefficients of the $\phi^2$ and $\phi^4$ terms in the Hamiltonian 
no longer  possess translational invariance. In order to absorb those terms,  we need 
to extend $(k^2 + t)$ and $u_1$ to inhomogeneous potentials.  In addition, the cubic 
interaction should be also included.  Thus the GL Hamiltonian $S(\equiv \be H)$ is generalized as   
\beq
 S[\phi; \rho] =  \sum_{l=1}^4 \frac{1}{l!} \int_{k_1,..., k_l} v_l (k_1,...,k_l) \phi(k_1) \cdots \phi(k_l), 
 \label{init}
\eeq
where $\rho$ represents all of the inhomogeneous potentials:
\beq
 \rho = \(v_1, v_2, v_3, v_4 \).  
\eeq

Starting with the Hamiltonian (\ref{init}),  we describe the RGT definitely.  
The higher and lower-momentum spaces are respectively defined as 
\beq
{\cal K_>} = \{ q : L^{-1} \Lambda < | q |  \leq \Lambda \}
\eeq
and 
 \beq
{\cal K_<} = \{ p : 0 < |p| \leq L^{-1} \Lambda \}
\eeq
with $L  > 1$. 
The Hamiltonian $S$ is decomposed into $S^<$ and $S^>$, where $S^<$ 
consists only of the lower momentum components of the field.  
The remaining part is denoted by $S^>$, i.e.,   
\beq
   S = S^{<} + S^{>}.  
 \label{decompose}
\eeq
Let us integrate over the higher-momentum components $\phi(q)$ in the partition function. Namely, 
\beqa
  Z &=& \int \prod_{k \in {\cal K}} d \phi(k) e^{-S} \nn\\
     &=& \int \prod_{p \in {\cal K_<}} d \phi(p) e^{-S^<} 
	           \int \prod_{q \in {\cal K_>}} d \phi(q) e^{-S^>}\nn\\
	 &=& \int \prod_{p \in {\cal K_<}} d \phi(p) e^{-S^< -\de {S}},  
\eeqa
where $\de S$ can be written as 
\beqa
  \de S &=& - \ln \int \prod_{k \in {\cal K_>}} d \phi(k) e^{-S^>}\nn\\
              &=& \sum_{l=1}^4 \frac{1}{l!} \int \de v_l (p_1,...,p_l) \phi(p_1) \cdots \phi(p_l) \nn\\
			  &&+ \mbox{ irrelevant terms}. 
\label{deS}
\eeqa
Next the scaling transformation is performed as
\beq
  \phi(p) = L^{\theta} \phi'(k), 
\eeq
where $k$ is related to $p$ by 
\beq
  k = L p.  
\eeq
Defining the new inhomogeneous potentials $v'_l$ by  
\beq
  v'_l \(k_1, ..., k_l\) \equiv L^{l\(\theta -d \)} \(v_l\(p_1,...,p_l\) + \de v_l\(p_1,...,p_l\) \), 
\label{ncc}
\eeq
the Hamiltonian turns back to the original form:
\beqa
  S^< + \de S &=&  \sum_{l=1}^4 \frac{1}{l!} \int_{k_1, ..., k_l} v'_l(k_1,...,k_l) \phi'(k_1)\cdots \phi'(k_l) \nn\\
 &=&S[\phi';\rho'], 
\eeqa
where $\rho' = (v_1', v_2', v_3', v_4' )$.  We then get the RGT for the potential $\rho \mapsto \rho'$.    
Furthermore, a correlation function transforms as 
\beq
 \bra \phi(p_1), \cdots, \phi(p_n) \ket_{S[\phi;\rho]} = L^{n \theta} \bra \phi(k_1), \cdots, 
 \phi(k_n) \ket_{S[\phi; \rho']}, 
\label{cf}
\eeq
where $\bra \cdot \ket_S$ means the thermal average using the Boltzmann weight $e^{-S}$.  

We proceed to the average over the random potentials.  It should be noted that Eqs. (\ref{de2}), (\ref{de4}) 
and (\ref{de3}) show that $v_2$, $v_3$ and $v_4$ come to have non-trivial correlations with each other. 
Therefore we can regard them as random potentials that  obey some  probability distribution 
as well as $v_1$.  We shall denote probability distribution by  $P[\rho]$.   
For example, if we take Eq. (\ref{H}) as the initial Hamiltonian, 
the corresponding distribution  $P_1$ is given by 
\begin{widetext}
\beqa
   P_1[\rho] 
   &=& P_0[v_1] \prod_{k_1,k_2,k_3,k_4}
   \de\(v_2\(k_1, k_2\) - \(k_1^2 + t \) \(2 \pi \)^d \de \(k_1 + k_2\) \) 
   \de\(v_3\(k_1, k_2, k_3\)\)\times \nn\\   
   &&\de\(v_4\(k_1, k_2, k_3, k_4\) - u_1 \(2 \pi \)^d \de\(\sum_{j=1}^4 k_j\) \).  
\eeqa
\end{widetext}

The RGT for the random potential $\rho \mapsto \rho'$ can be pushed into 
change of $P$ by the following rule:
\beq
  \int F[\rho'] P[\rho] {\cal D} \rho = \int F[\rho] P'[\rho] {\cal D \rho}, 
\label{rgt-P}
\eeq
where $F$ is an arbitrary functional of $\rho$ and 
\beq
  {\cal D}\rho \equiv \prod_{l=1}^4 \prod_{k_1, ..., k_l} d v_l(k_1,...,k_l).  
\eeq
Namely, $P'$ is formally written as\cite{hl}
\beq
  P'[\rho'] = P[\rho(\rho')] \left|\frac{\del \rho(\rho')}{\del \rho'} \right|.  
\label{p'}
\eeq
Eq.  (\ref{rgt-P}) defines RGT $P \mapsto P'$,  keeping the Hamiltonian invariant.  For instance, 
using Eq. (\ref{cf}),  the random-potential average of the correlation function becomes
\beqa
   && \int P[\rho] \bra \phi(p_1), \cdots, \phi(p_n) \ket_{S[\phi;\rho]} {\cal D} \rho \nn\\
   &=& 
  \int P'[\rho] L^{n \theta} \bra \phi(k_1), \cdots, \phi(k_n) \ket_{S[\phi; \rho]}{\cal D} \rho.
\eeqa

In the practical computation of  the random-potential average, 
we give cumulants among random potentials instead of giving 
explicit form to the probability distribution.  First, as in the original theory,  
we put
\beqa
  &&\rnd{v_1(k_1) ; v_1(k_2)} = \Delta \(2\pi\)^d \de(k_1 + k_2) \nn\\
  && \rnd{v_2(k_1, k_2)} = (k_1^2 + t) \(2\pi\)^d \de(k_1 + k_2) \nn\\
  &&\rnd{v_4(k_1, k_2,k_3, k_4)} = u_1 \(2\pi\)^d \de\(\sum_{j=1}^4 k_j \),
\label{rnd0}
\eeqa
where $\rnd{\ \cdot \ }$ means to take the average over the random variables with 
 distribution $P[\rho]$. 
The semicolon in the bracket means the cumulant product: 
e.g.,  $\rnd{X;Y} \equiv \rnd{XY} - \rnd{X} \rnd{Y}$.  
We assume that there is no long-range correlation between random variables.  In this assumption, 
the following non-vanishing cumulants are added to Eq. (\ref{rnd0}): 
\beqa
 &&\left(
         \begin{array}{l}
		   \rnd{v_1(k_1); v_3(k_2, k_3, k_4)} \\
		   \rnd{v_2(k_1, k_2); v_2(k_3, k_4)} \\
		   \rnd{v_2(k_1, k_2); v_1(k_3); v_1(k_4)} \\
		   \rnd{v_1(k_1);v_1(k_2);v_1(k_3);v_1(k_4)} 
		 \end{array}
 \right) \nn\\
&=&
 \left(
          \begin{array}{c}
		    u_2 \\
			u_3 \\
			u_4 \\
			u_5
		  \end{array}
  \right)
    \(2 \pi\)^d \de \(\sum_{j=1}^4 k_j \). 
\label{rnd}
\eeqa
Note that  we respect the symmetry 
\beq
 v_1 \rightarrow -v_1, \ \ \phi \rightarrow -\phi, 
\eeq
so that 
\beq
    \rnd{v_1(k_1)} = \rnd{v_3(k_1, k_2, k_3)} = 0.  
  \eeq
The parameters $u_j (j=1, ..., 5)$ have the scaling dimension $4-d$ 
(measured by the inverse length). 
Cumulants other than Eqs. (\ref{rnd0}) and (\ref{rnd}) are ignored since they 
are associated with higher-dimensional parameters\footnote{
However, there are apparent higher-dimensional parameters that effectively have  the 
same scaling dimensions as $u_{i}$\cite{fe}.  See Sec.  \ref{observation}.
}. 
Probability distribution $P$ is characterized by  parameters $(t, \Delta, u_1, ..., u_5) \equiv \mu$.  
Transformed distribution $P'$ is  similarly characterized by $(t', \Delta', u'_1, ..., u'_5) \equiv \mu'$.
Namely, $\mu'$ is obtained taking the random-potential
 average of Eqs. (\ref{rnd0}) and (\ref{rnd}) in use of 
$P'[\rho]$ instead of $P[\rho]$.  

Transformation $\mu \mapsto \mu'$ is analyzed using the following formula:
\beqa
  &&\rndp{v_{l_1}(k^1_1,...,k^1_{l_1}); \cdots ;v_{l_n}(k^n_1,...,k^n_{l_n})} \nn\\
 = &&\rnd{v'_{l_1}(k^1_1,...,k^1_{l_1}); \cdots ;v'_{l_n}(k^n_1,...,k^n_{l_n})}, 
\label{pp'} 
\eeqa
which is easily checked from Eq. (\ref{rgt-P}).  We apply it to the non-vanishing cumulants. 
By definition, the left-hand side can be written in  $\mu'$.  On the other hand,  
if the transformed potentials are described by the original ones employing 
Eq. (\ref{ncc}), the right-hand side can be  evaluated in terms of $\mu$.  
In this way, $\mu'$ is expressed in terms of $\mu$.   
 
\section{Perturbative renormalization group}
\label{pert}
\subsection{Diagrammatic expansion}
\label{diag}
In this section we obtain transformation $\mu \mapsto \mu'$ by perturbation.  
We consider all $u_j$s as small parameters and express 
$\mu'$ up to ${\rm O}(u_i u_j)$. On the other hand,  we do not regard $t$ and 
$\Delta$ as small since they are relevant parameters of  the dimension 2. 

We can formulate the perturbative expansion by adopting  the random-potential 
average of $v_2(q_1, q_2)$ 
and the linear term as the unperturbed Hamiltonian $S_0$:
\beqa
     S_0 &= &\frac{1}{2}\int_{k_1,k_2} \rnd{v_2(k_1, k_2)} \phi(k_1) \phi(k_2)+ \int_k v_1(-k) \phi(k) \nn\\
	     &= & \frac{1}{2}\int_{k} (k^2 + t) \varphi(k) \varphi(-k) - \frac{1}{2} \int_k  v_1(-k) G(k) v_1(k), 
\nn\\
\eeqa
where 
\beqa
\varphi(k) &\equiv& \phi(k) + G(k) v_1(k)\nn\\
G(k) &=& \frac{1}{k^2 + t}. 
\eeqa
The remaining terms are denoted by  $V$ and treated as the perturbative Hamiltonian. 
Let 
\beq
  V = \sum_{j=2}^4 V_j,  
\eeq
where 
\beqa
    V_2 &=& \frac{1}{2} \int_{k_1,k_2} \( v_2 (k_1, k_2) - \rnd{v_2(k_1,k_2)} \)\phi(k_1)\phi(k_2) \nn\\
	V_j  &=& \frac{1}{j!} \int_{k_1,...,k_j} v_j (k_1, ..., k_j) \phi(k_1) \cdots \phi(k_j), 
	\nn\\
	&&(j = 3, 4).  
\eeqa

As we have performed in Eq. (\ref{decompose}), we 
divide $S_0$ and  $V$ into 
\beqa
  S_0 &=& S_0^> + S_0^< \nn\\
  V &=& V^> + V^<
\eeqa
respectively.  
The free part of the partition function is denoted by 
\beq
  Z_0 = \int \prod_{q \in \kh} d\phi(q) e^{-S^>_0}.  
\eeq
Writing 
\beq
  \coarse{X} \equiv \frac{1}{Z_0} \int \prod_{q \in \kh} d\phi(q) X e^{-S_0^>}, 
\eeq
$\de S$ in Eq. (\ref{deS})  is expanded as 
\beqa
  -\de S &=& \log Z_0 + \log \coarse{e^{-V^>}}\nn\\
&=& \log Z_0  - 
  \coarse{{V^>}} + 
  \frac{1}{2}\coarse{V^>; V^>}  
  \nn\\
  &&-
  \frac{1}{3!} \coarse{{V^>;V^>; V^>}} 
  + {\rm O}({V^>}^4),  
\label{exp}
\eeqa
where the semicolon represents the cumulant with respect to the thermal average $\coarse{\ }$.  
Note that, in practical computation, $\phi(q)$ in $V^>$ is understood as 
$\varphi(q) - G(q)v_1(q)$ and the Wick theorem is applied to $\varphi(q)$. 
Comparing Eqs.(\ref{deS}) and (\ref{exp}),  we can obtain the perturbative expansion of 
$\de v_l (p_1,...,p_l)$, which leads to transformed potential $v'_l$ 
defined in Eq. (\ref{ncc}).  Inserting the resultant $v'_l$ into the right-hand side of Eq. (\ref{pp'}), 
we can express the transformed parameters in terms of the original ones. In order to proceed 
with this program efficiently,  we will use the diagrammatic technique.  

\begin{figure}
\begin{center}
\setlength{\unitlength}{1mm}
\begin{picture}(80, 20)(0,0)
        \put(0,0){ 
		\includegraphics[width=80mm]{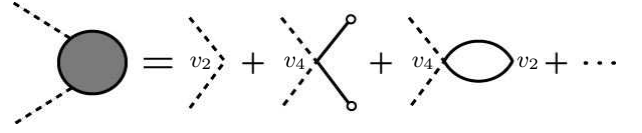}
		}
				\put(24.5,7.5){$v_2$}
				\put(37,7.5){$v_4$}
				\put(54,7.5){$v_4$}
				\put(68,7.5){$v_2$}
\end{picture}
\end{center}
\caption{Corrected potential $v'_2$.   An internal line 
carrying higher momentum is represented by a solid line, 
while an external one by a dashed line.  An open circle stands for $v_1(q)$. }
\label{fig_vertex}
\end{figure}

Let us depict $v'_l$ by a shaded circle with $l$ dashed lines.  
FIG. \ref{fig_vertex} exemplifies the case of $l=2$.  
%
The shaded circle contains $\de v_l$, which is expanded by {\em connected} diagrams, 
 according to Eq. (\ref{exp}).
The random potential $v_l(k_1, ..., k_l)$ makes an $l$-point vertex. 
In particular,  $v_1(q) (q \in \kh)$ is expressed by an open circle.  
It should be noted that  total momentum is not conserved 
at the vertex $v_l(k_1, ..., k_l)$.  
 
\begin{figure}
\begin{center}
\setlength{\unitlength}{1mm}
 \begin{picture}(60, 50)(0,0)
        \put(0,0){ 
		\includegraphics[width=60mm]{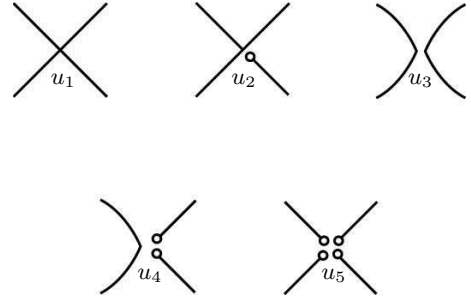}
		}
		\put(6.0,28){$u_1$}
		\put(17.5,2){$u_4$}
		\put(30,28){$u_2$}
		\put(53.5,28){$u_3$}
		\put(42,2){$u_5$}
\end{picture}
\end{center}
\caption{four-point vertices}
\label{fig_vertices}
\end{figure}

When  we take the random-potential
 average, non-vanishing cumulants $\rnd{v_4}$, $\rnd{v_3; v_1}$, 
 $\rnd{v_2; v_2}$, $\rnd{v_2; v_1; v_1}$ and $\rnd{v_1; v_1;v_1; v_1}$ 
 can be graphically represented as four-point vertices at which the total 
 momenta,  carried by the four lines,  are conserved due to the delta 
 function in Eq. (\ref{rnd}). See FIG. \ref{fig_vertices}.  
%
%

\begin{figure}
\begin{center}
\setlength{\unitlength}{1mm}
\begin{picture}(50, 20)(0,0)
        \put(0,0){ 
		\includegraphics[width=50mm]{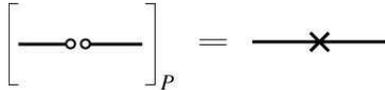}}
\end{picture}
\end{center}
\caption{Merging two open circles}
\label{fig_cross}
\end{figure}

Further, the cumulant $\rnd{v_1; v_1}$ tells us that two internal lines 
ended with the open circles are merged by the random-potential
 average and produce $\Delta$.  We here depict it by a cross, $\times$, 
as in FIG. \ref{fig_cross}.  The cross reminds us that the two internal lines are not connected with 
respect to the thermal average.  

%
\begin{figure}[h]
\begin{center}
\setlength{\unitlength}{1mm}
\begin{picture}(80, 105)(0,0)
		\includegraphics[width=80mm]{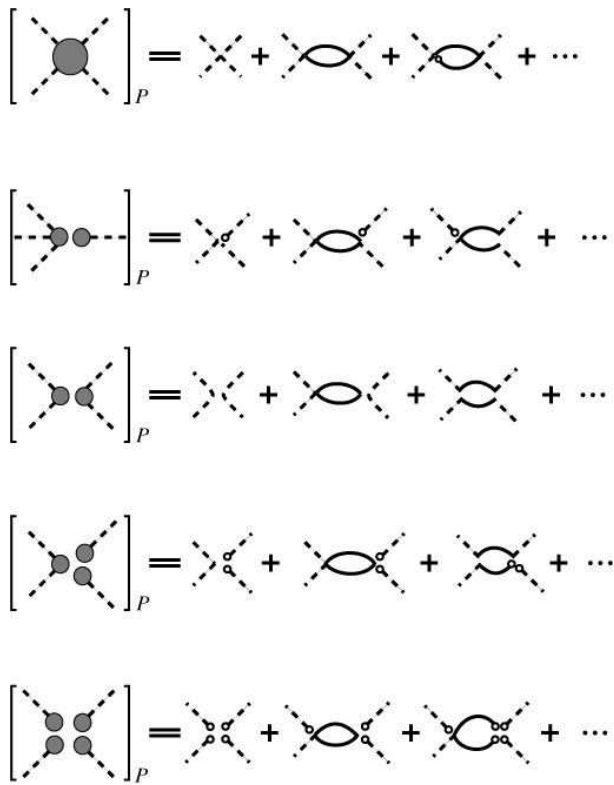}
\end{picture}
\end{center}
\caption{Perturbative expansion for cumulants contributing $u'_j$ $(j=1,...,5)$}
\label{fig_pert}
\end{figure}

The right-hand side of Eq. (\ref{pp'}) is graphically represented 
as the left-hand side of FIG. \ref{fig_pert}.  
When we expand the shaded circles in FIG. \ref{fig_pert}, 
we get a summation of the diagrams that contain 
several connected components,  as shown in the right-hand side of the figure.  
%
%
If we want to know  $u'_i$ in ${\rm O}(u_j u_k)$, we need to consider the one-loop diagrams made of the vertices in FIG. \ref{fig_vertices} and $\times$, including the numerical factors of those diagrams. Note that
a one-loop diagram that contains a connected component without external legs does not appear in the right-hand side of FIG. \ref{fig_pert},  because the shaded circles in the left-hand side consist of only connected diagrams.  In particular,  
it is ruled out that there are  two or more
$\times$s on an internal line. 

In order to write down all one-loop diagrams in the right-hand side of FIG. \ref{fig_pert}, 
we must first consider  all the admissible diagrams without $\times$,  and then put $\times$ 
on as many internal lines as possible.  After we obtain the set of all one-loop diagrams, 
we have to compute their numerical factors.  Although the procedure is fulfilled 
 in  appendix \ref{recursion},  here we present the following examples.

\subsection{Example}
\label{exam}
\begin{figure}[h]
\begin{center}
\setlength{\unitlength}{1mm}
\begin{picture}(60, 40)(0,0)
       \put(0,0){ 
		\includegraphics[width=60mm]{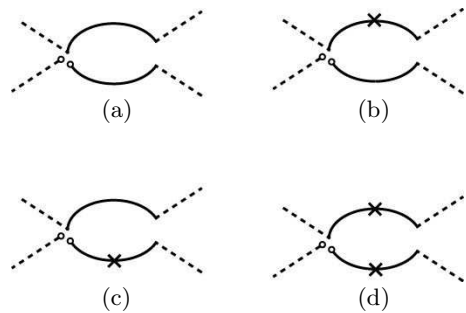}
		}
		\put(13,22){(a)}
		\put(47,22){(b)}
		\put(13,-3){(c)}
		\put(47,-3){(d)}
\end{picture}
\end{center}
\caption{Diagrams generated from (a) by putting $\times$.}
\label{fig_ce}
\end{figure}

Let us consider FIG. \ref{fig_ce}(a).    
Putting $\times$ on the internal lines, (b), (c), and (d) in the figure are generated.   
However,  (c) and (d) are excluded because 
they contain  connected components without external lines.  
Counting the external lines of the connected components in 
FIG. \ref{fig_ce}(a), we find that it makes a contribution to $u'_4$,  
while FIG.  \ref{fig_ce}(b) contributes to $u'_5$.
%
%
%
First we compute the contribution from FIG. \ref{fig_ce}(a) to      
\beqa
 u_4' \(2\pi\)^d \de \(\sum_{k=1}^4 k_i\) =  
  \rnd{v'_2(k_1, k_2), v'_1(k_3); v'_1(k_4)}. 
\label{u4'}
\eeqa
Using Eq. (\ref{ncc}), FIG. \ref{fig_ce}(a) comes from 
\beq
  L^{4(\theta -d)} \rnd{\de v_2(p_1, p_2); \de v_1(p_3) ; v_1(p_4)} 
  + \(p_3 \leftrightarrow p_4 \).  
\label{u4}
\eeq
The second term arises because 
the roles of $v_1'(k_3)$ and $v_1'(k_4)$ can interchange.  The two terms in the right-hand side of 
Eq. (\ref{u4}) equally contribute to $u_4$.  
In general, if a cumulant contains 
multiple potentials of the same kind, the number of ways of 
 interchanging should be counted. 
We shall denote the number with $n_R$.  As for the present case, 
\beq
  n_R = 2. 
\eeq

Now we compute the first term.  Defining 
\beqa
  &&V_l^{(m)}\(p_1, ..., p_m\) \equiv \frac{1}{l!} {l \choose m } \times 
  \nn\\
  && \int_{q_1,..., q_{l-m}}^> \! v_l   \(p_1, ..., p_m, q_1,.., q_{l-m} \) 
  \prod_{i=1}^{l-m}\phi\(q_i\), 
\label{vlm}
\eeqa 
we show explicitly the  terms in $\de v_2(p_1, p_2)$ necessary for making FIG. \ref{fig_ce}(a) as  
\beqa
   \de v_2(p_1, p_2) &=& 
   - 2! \cdot \frac{1}{2!}\coarse{V_2^{(1)}(p_1); V_2^{(1)}(p_2)} + \cdots \nn\\
  &\equiv& a_1 \int^>_{q_1} v_2(p_1, q_1) v_2(p_2, -q_1) G(q_1) \nn\\
  &&+ \cdots .  
\label{dev2}
\eeqa
Here the dots in the right-hand side indicate terms not employed for FIG. \ref{fig_ce}(a).  
 In the first equality, the factor $2!$ originates from the normalization 
 of $\de v_2(p_1, p_2)$,  while $1/2!$ comes from the expansion of $e^{-V_2}$. 
 The minus sign is so put because it is a second-order perturbation term.  
Performing the thermal average,  we can easily check that 
 \beq
   a_1 = -1. 
 \eeq
Similarly, 
 \beqa
   \de v_1(p) &=& - \coarse{V_2^{(1)}(p)} + \cdots \nn\\
   &=& a_2 \int^>_{q_2} v_2(p_3, q_2) v_1(-q_2) G(q_2)
 \label{dev1}
 \eeqa
 with
 \beq
 a_2 = -1.
 \eeq
The factors $a_1$ and $a_2$ are associated with the connected components 
of the diagram we have computed. We shall denote the product of 
them with $n_A$, i.e., 
\beq
  n_A  = a_1 a_2 = 1. 
\eeq

Next we go to the random-potential average. 
Inserting Eqs.(\ref{dev2}) and (\ref{dev1}) into the 
first term of Eq. (\ref{u4}), the random-potential average becomes 
\begin{widetext}
\beqa
 && L^{4(\theta -d)}\int^>_{q_1,q_2}G(q_1) G(q_2)
\rnd{v_2(p_1, q_1) v_2(p_2, -q_1); v_2(p_3, q_2) v_1(-q_2); v_1(p_4)}\nn\\
 &=&  L^{4(\theta -d)} \int^>_{q_1,q_2}G(q_1) G(q_2)
\rnd{v_2(p_1, q_1 ); v_2(p_3, q_2)}  \rnd{v_2(p_2, -q_1) ; v_1(-q_2); v_1(p_4)} 
\nn\\
&&+ 
 \(v_2\(p_1, q_1\) \leftrightarrow v_2\(p_2, -q_1\) \)\nn\\
 &=& L^{4\theta -3d} u_3 u_4 \(2\pi \)^{d} \de \(\sum_{i=1}^4 k_i \)\int^>_q \(G(q)G(p_2+p_4-q) + G(q)G(p_1+p_4+q) \).
\label{rndav}
\eeqa 
\end{widetext}
Note that there are two ways of contraction with respect to the random-potential 
 average in the first equality. 
We denote the number of ways of the  contraction  by $n_C$. That is, 
\beq
  n_C = 2
\eeq
in the case of FIG. \ref{fig_ce}(a).

Thus the correction to $u_4$ by way of FIG. \ref{fig_ce}(a) is read from Eq. (\ref{rndav}),  combining
 with the second term of Eq. (\ref{u4}).  Letting $p_i = 0$ $(i=1,...,4)$ in $G$, we find that the numerical 
factor $n_F$ associated with the diagram is 
\beq
  n_F = n_A n_R n_C = 4.  
\eeq
Therefore the correction to $u_4$ is 
\beq
  n_F L^{4\theta -3d} u_3 u_4 \int^>_q G^2(q)  = 4L^{4\theta -3d} u_3 u_4 \int^>_q G^2(q).  
\eeq

Let us turn to FIG. \ref{fig_ce}(b), which  affects $u'_5\propto 
\rnd{v'_1(p_1);v'_1(p_2);v'_1(p_3);v'_1(p_4) }$.   
For creating this diagram,  we need three $ \coarse{V_2^{(1)}(p)}$ and one $v_1(p)$. Obviously, 
\beq
  n_R = 4. 
\eeq
The numerical factor 
associated with the connected component was already computed in Eq. (\ref{dev1}). 
 Their product  becomes  
\beq
   n_A = \(-1\)^3 \times 1 = -1. 
\eeq
Since the number of contractions is equal to the  number of  ways of contracting  
$v_2(p, q)$ and $v_1(p)$, 
\beq
 n_C = 3. 
\eeq
Thus the contribution to $u'_5$ of FIG. \ref{fig_ce}(b) is 
\beq
  n_F \Delta u_2 u_3 L^{4\theta -3 d}\int_q^>  G(q)^3,   
\eeq
with 
\beq
  n_{F} = n_{R}n_{C}n_{A} = -12. 
\eeq

\subsection{Results}
After writing all one-loop diagrams appearing in the right-hand side of FIG. \ref{fig_pert}, 
we can obtain the recursion equation in ${\rm  O}\(u_i u_j\)$.   
The coefficients $n_A$, $n_C$, and $n_R$ associated with 
each diagrams are presented in  appendix \ref{recursion}.  
Here we present the result. 

\beqa
  &&u'_j = L^{4 \theta -3 d} \( u_j + \de u_j \), \ \ \(j = 1, ..., 5\) \nn\\
  &&t' = L^{2 \theta -d}  \( t + \de t\)\nn\\
  &&\Delta' = L^{2 \theta -d} \( \Delta + \de \Delta \)
\label{cor2}
\eeqa
with 
 \begin{widetext}
\beqa
 \de u_1 &=&  -\( 3  {\cal A}_3 \Delta + \frac{3}{2} {\cal A}_2 \)u_1^2+ 6 {\cal A}_2 u_1 \tilde{u}_3   \nn\\
 \de u_2 &=&  - 3   {\cal A}_3 \Delta u_1\tilde{u}_3  -\frac{3}{2}  {\cal A}_2 u_1 u_2
 + 6 {\cal A}_2 u_2 \tilde{u}_3 - 3 {\cal A}_2 u^2_2 + 3 {\cal A}_2 u_1 u_4  \nn\\
 \de u_3 &=& \frac{1}{2}  {\cal A}_4 \Delta^2 u_1^2 -  \(2{\cal A}_3  \Delta+ 
  {\cal A}_2 \) u_1 \tilde{u}_3 +{\cal A}_2 u_1u_2+{\cal A}_2  u_1 u_4 +
4 {\cal A}_2  \tilde{u}_3^2 - 3{\cal A}_2 u_2^2-6{\cal A}_2 u_2 \tilde{u}_3
 \nn\\
 \de u_4 &=& {\cal A}_4 \Delta^2 u_1 \tilde{u}_3   -\(3 {\cal A}_3\Delta+\frac{1}{2}  {\cal A}_2\) u_1 u_4   
 - 4{\cal A}_3 \Delta \tilde{u}_3^2  
 + \frac{1}{2} {\cal A}_2 u_1 u_5   - {\cal A}_2 u_2 \tilde{u}_3 + 8  {\cal A}_2\tilde{u}_3 u_4 
  \nn\\
 \de u_5 &=& 6 {\cal A}_4\Delta^2 \tilde{u}_3^2 - 36   {\cal A}_3 \Delta \tilde{u}_3 u_4
              -6 {\cal A}_2u_2 u_4 + 6{\cal A}_2 \tilde{u}_3 u_5 + 24{\cal A}_2 u_4^2, 
\label{cor1}
 \eeqa
 \end{widetext}
where we have defined 
\beq
  {\cal A}_n \equiv \int_q^> G(q)^n
\eeq 
and 
\beq
 \tilde{u}_3 \equiv u_2 + u_3.  
\label{tildeu3}
\eeq

 Similarly, $\de t$ and $\de \Delta$ are given as 
 \begin{widetext}
 \beqa
\de t &=& \frac{1}{2}\({\cal A}_2 \Delta + {\cal A}_1\)  u_1 
        - {\cal A}_1  \tilde{u}_3 
        \nn\\
        &&+\( \(-\frac{1}{2}{\cal A}_2 {\cal A}_3 
   		- \frac{1}{2}  {\cal B}_2\) \Delta^2 
		-\(\frac{1}{2} {\cal A}_1 {\cal A}_3 +\frac{1}{2} {\cal B}_1+\frac{1}{4}{\cal A}_2^2 \)\Delta
		  -\frac{1}{4}{\cal A}_1 {\cal A}_2
		 -\frac{1}{6}{\cal B}_0 \) u_1^2
\nn\\
        &&+ \frac{1}{2}{\cal A}_1 {\cal A}_2 u_1 u_2 +\( \({\cal A}_1 {\cal A}_3 
		+ {\cal A}_2^2 +  2{\cal B}_1\) \Delta
		+ {\cal A}_1 {\cal A}_2+{\cal B}_0\) u_1 \tilde{u}_3 
\nn\\
        &&
		- \({\cal A}_1 {\cal A}_2+ {\cal B}_0\)u_1 u_4
		-  \({\cal A}_1 {\cal A}_2+{\cal B}_0\)\tilde{u}^2_3
\nn\\ 
\de \Delta &=& {\cal A}_2 \Delta \tilde{u}_3  +  u_2 {\cal A}_1  - 2 u_4 {\cal A}_1 
        +\frac{1}{6} {\cal B}_3 \Delta^3u_1^2 
		-\(\frac{1}{2}{\cal A}_2^2\Delta+\frac{1}{2}{\cal A}_1{\cal A}_2
		+\frac{1}{3}{\cal B}_0\)u_1 u_2
\nn\\
 &&	-\(\({\cal A}_2 {\cal A}_3+2{\cal B}_2\) \Delta^2 
		+ \({\cal A}_1{\cal A}_3+{\cal B}_1\)\Delta\)u_1 \tilde{u}_3
 \nn\\
 &&      +  \( \({\cal A}_2^2+ 3{\cal B}_1\)\Delta+{\cal A}_1{\cal A}_2+{\cal B}_0\) u_1 u_4
            -\frac{1}{3}{\cal B}_0u_1u_5-{\cal B}_0u_2^2
\nn\\
 &&+ \(2 {\cal A}_1{\cal A}_2 +2{\cal B}_0\) u_2\tilde{u}_3
     + \( 3{\cal B}_1+  {\cal A}_2^2+ 2  {\cal A}_1{\cal A}_3\)\Delta \tilde{u}_3^2
      -\(4 {\cal A}_1{\cal A}_2+ 4 {\cal B}_0\)\tilde{u}_3u_4, 
\label{cor3}
\eeqa
\end{widetext}
where 
\beqa
 && {\cal B}_0 \equiv \int^>_{q_1,q_2} G(q_1) G(q_2) G(q_1+q_2) \nn\\
&& {\cal B}_1 \equiv \int^>_{q_1,q_2} G(q_1)^2 G(q_2) G(q_1+q_2) \nn\\ 
 && {\cal B}_2 \equiv \int^>_{q_1,q_2} G(q_1)^2 G(q_2)^2 G(q_1+q_2)\nn\\ 
&& {\cal B}_3 \equiv \int^>_{q_1,q_2} G(q_1)^2 G(q_2)^2 G(q_1+q_2)^2. 
\eeqa

\subsection{Comparison to the replica method}
\label{comp}
We compare the recursion relations shown above 
to those obtained by the replica method\cite{bd,sak}, 
where the perturbative Hamiltonian is 
\beqa
 S_{\rm int}^{\rm rep} &=& 
\frac{u^{\rm rep}_1}{4!} \sum_{\al=1}^n  \phi_\al^4  + 
\frac{u^{\rm rep}_2}{3!} \sum_{\al,\be=1^n} \phi_\al^3 \phi_\be 
\nn\\
&&+\frac{u^{\rm rep}_3}{8} \sum_{\al,\be=1}^n \phi_\al^2 \phi_\be^2
+\frac{u^{\rm rep}_4}{4} \sum_{\al,\be,\ga=1}^n \phi_\al^2 \phi_\be \phi_\ga 
\nn\\
&& +\frac{u^{\rm rep}_5}{4!} \sum_{\al,\be,\ga,\de=1}^n 
\phi_\al \phi_\be \phi_\ga \phi_\de. 
\label{srep}
\eeqa
Br\'ezin and de Dominicis proposed the two limiting procedures listed below:
\begin{description}
 \item{(A)} $u^{\rm rep}_1, ..., u^{\rm rep}_5$ are fixed, then take $ n\rightarrow 0$.
 \item{(B)} $u^{\rm rep}_1, n u^{\rm rep}_2, n u^{\rm rep}_3, n^2u^{\rm rep}_4, n^3 u^{\rm rep}_5$
                are fixed and take $n \rightarrow 0$.   
\end{description}
We can check that the recursion equations (\ref{cor2}) - (\ref{tildeu3}) 
are consistent with the procedure (A) by the following identification
\beqa
  &u_i = u^{\rm rep}_i,  &\mbox{ for $i=1, 4$}\nn\\
  &u_j  = - u^{\rm rep}_j,  &\mbox{ for $j = 2,3,5$}. 
  \label{identify}
\eeqa
More precisely, we pick the correction terms proportional to ${\cal A}_3$ in Eq. (\ref{cor3}), 
 which 
is estimated as $\log L$ in $d=6-\epsilon$. Putting $L = e^{\de l}$ and taking the first order 
of $\de l$, we get the beta functions in Refs.\cite{bd,sak}.   
This is an expected result,  because we can obtain the replicated Hamiltonian $S'_{\rm rep}$
from the cumulant expansion
\beq
  S'_{\rm rep} = \sum_{j=1}^{\infty}\frac{\(-1\)^{j}}{j!}
  \rnd{\sum_{\al}S[\phi_{\al}; \rho] ; 
  \cdots ; \sum_{\be}S[\phi_{\be}; \rho] }
\eeq
with use of 
 Eqs.(\ref{rnd0}) and (\ref{rnd}).  We find that $S'_{\rm rep}$ with 
Eq. (\ref{identify}) gives Eq. (\ref{srep}).

The procedure (B) contains some diagrams  of ${\rm O}(n)$.   These diagrams never 
appear by  way of the present method,  because those contain a connected component 
without external lines. 



\section{Expansion parameters}
\label{redef}
\subsection{Preliminaries}
Since all $u_j$s have the dimension $4-d$, they are apparently irrelevant when 
$d>4$.  However,  coefficients of the perturbation series in 
$\{u_j\}$ are expressed  as polynomial of $\Delta$ having the dimension 2, which can change 
the relevance of $u_j$s.  That is,  since $\Delta$ is relevant for all dimensions,  
a term containing it can become larger as 
repeating the RGT.  Hence,  it is important to 
 explore how $\Delta$ appears in the perturbation series.  
 
\begin{figure}[h]
\begin{center}
\setlength{\unitlength}{1mm}
 \begin{picture}(60, 25)(0,0)
        \put(0,10){ 
		\includegraphics[width=55mm]{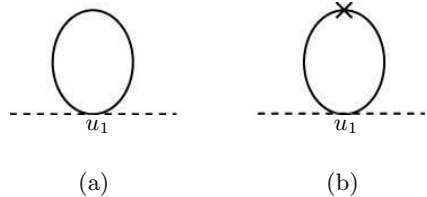}
		}
       \put(11,8){$u_1$}
       \put(44,8){$u_1$}
       \put(10,0){(a)}
       \put(43,0){(b)}
\end{picture}
\end{center}
\caption{Leading correction from $u_1$ to $\de t$. }
\label{fig_tadpole1}
\end{figure}

To make our argument clear, let us consider $\de t$ as an  example.   
The coefficient of the first order 
in  $u_1$, which appears in the first term of $\de t$ in Eq. (\ref{cor3}), 
is depicted by the diagrams in FIG. \ref{fig_tadpole1}.  The internal line of (a) in the figure 
carries $G(q)$, which  behaves as $1/q^2$ at the criticality, while the internal line of (b) brings 
$\Delta G(q)^2\sim \Delta/q^4$. 
%
%
Namely, the dominant contribution is proportional to $\Delta u_1$,  as 
 pointed out in the previous literatures\cite{aim,g}.  
We  then introduce  
\beq
  g_1 \equiv \Delta u_1.  
\eeq
as one of expansion parameters.    
It has the scaling dimension $6-d$, 
 which implies that the upper critical dimension is six.  Therefore we cannot ignore 
 $u_1$ in $d\leq 6$,  even though $u_1$ itself is irrelevant when $4<d$.  

In addition, 
we define
\beq
  g_0 \equiv \Delta^{-1}. 
\eeq
FIG. \ref{fig_tadpole1}(a) is then proportional to $u_1 = g_0 g_1$. Generally, if a diagram includes 
the irrelevant parameter 
$g_0$,  its contribution is less relevant.  

Next we discuss how $\Delta$ couples with  the other parameters $u_2, ..., u_5$. 
Let us look at the first-order terms in $u_2$ and $u_3$ respectively, 
which  appear in the second term of $\de t$ in Eq. (\ref{cor3}).  
(Note: we have defined $\tilde{u}_3 \equiv u_2 + u_3$.)
%
\begin{figure}[h]
\begin{center}
\setlength{\unitlength}{1mm}
 \begin{picture}(60,53)(0,0)
        \put(0,8){ 
		\includegraphics[width=55mm]{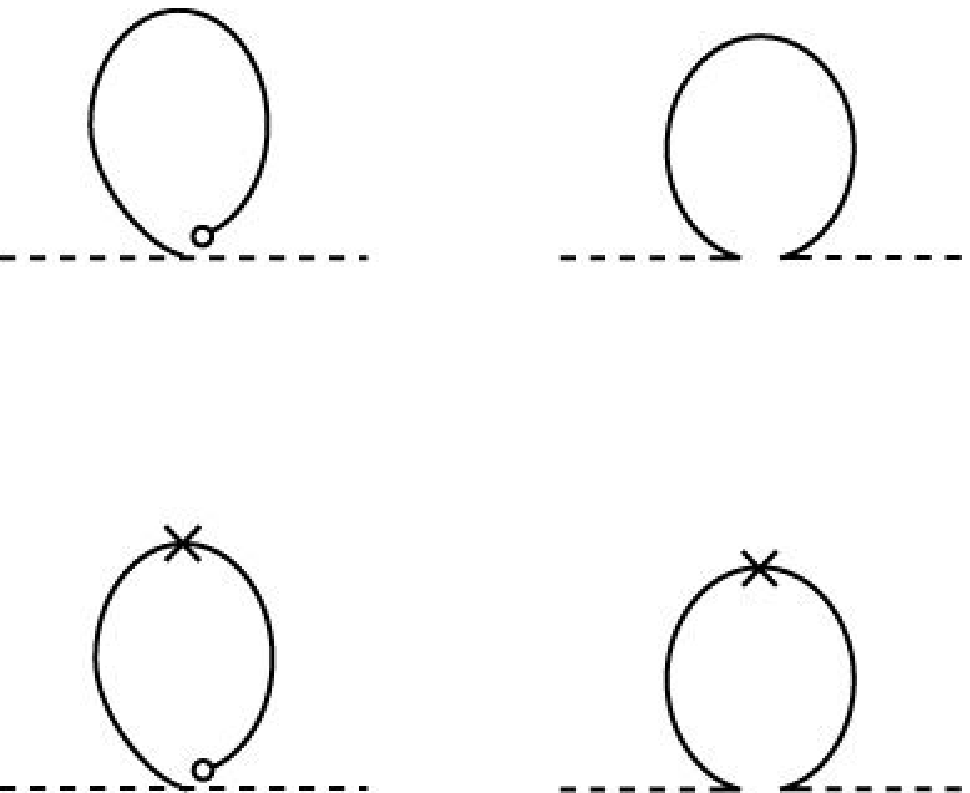}
		}
      \put(10,35){$u_2$}
      \put(43,35){$u_3$}
      \put(10,5){$u_2$}
      \put(43,5){$u_3$}
      \put(10,30){(a)}
      \put(42,30){(b)}
      \put(10,0){(c)}
      \put(42,0){(d)}
\end{picture}
\end{center}
\caption{Diagrams in (a) and (b) give leading correction to $t$ by $u_2$ and $u_3$ 
respectively.  
However,  diagrams in (c) and (d) do not contribute to $\de t$,  because
 they both have two connected components before taking the random-field average. }
\label{fig_tadpole2}
\end{figure}
%
It turns out that their coefficients do not carry $\Delta$ because putting $\times$ on 
the internal line
 produces  disconnected components as shown in FIG. \ref{fig_tadpole2}(c) and (d), 
 so that (c) and (d) do not appear in the perturbation series. 
 Thus no  $\Delta$ 
 associates with  $u_2$ and $u_3$,  so that we choose 
 \beqa
   &&g_2 \equiv \Delta^0  u_2  =  u_2
   \nn\\
   &&g_3 \equiv \Delta^0 u_3  = u_3 
 \eeqa
as  expansion parameters.  They have the scaling dimension  $4-d$.  

\begin{figure}[h]
\begin{center}
\setlength{\unitlength}{1mm}
 \begin{picture}(60,15)(0,0)
        \put(0,0){ 
		\includegraphics[width=55mm]{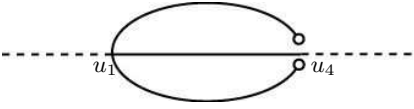}
		}
       \put(13,4){$u_1$}
       \put(42,4){$u_4$}
\end{picture}
\end{center}
\caption{One of the leading contributions from $u_4$ to $\de t$. }
\label{fig_u1u4}
\end{figure}

Let us turn to the parameter $u_4$. It  contributes to $\de t$ by the combination
$u_1 u_4$ in the lowest order.  
See FIG. \ref{fig_u1u4}.  Here $\times$ is forbidden,  for  the same reason it is in  
FIG. \ref{fig_tadpole2}(c). 
%
%

\noindent
Therefore,  this contribution is proportional to 
$u_1 u_4 = g_1 \( \Delta^{-1} u_4 \)$. Then we add 
\beq
  g_4 \equiv \Delta^{-1} u_4
\eeq
to the expansion parameters.  It has the 
scaling dimension $2-d$. 

\begin{figure}[h]
\begin{center}
\setlength{\unitlength}{1mm}
 \begin{picture}(60,13)(0,0)
        \put(0,0){ 
		\includegraphics[width=55mm]{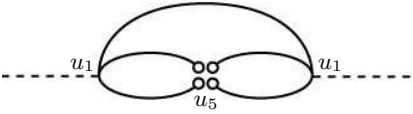}
		}
       \put(10,4){$u_1$}
       \put(26.5,-1){$u_5$}
       \put(43,4){$u_1$}
\end{picture}
\end{center}
\caption{One of the leading contributions from $u_5$ to $\de t$. }
\label{fig_u1u5}
\end{figure}

As for $u_5$, the leading correction to $\de t$ comes from the order in  
$u_1^2 u_5 = g_1^2 \(\Delta^{-2} u_5\)$, 
as we depicted in FIG. \ref{fig_u1u5}. 
%
%

\noindent
Hence we define 
\beq
  g_5 = \Delta^{-2} u_5, 
\eeq
which has the dimension $-d$.  

In this way, we can express the perturbation series of 
$\de t$  in terms  of $g_0, ..., g_5$.   
   Further,
we can rewrite the recursion equations for 
$\Delta$ and $u_i$ $(1=1, ..., 5)$ into those for $g_\mu$ $(\mu = 0, ..., 5)$.  
It is easily confirmed  from the explicit form of Eqs.(\ref{cor1}) and (\ref{cor3}) 
that no $\Delta$ appears in these series at least in the second order 
in $\{g_{\mu}\}$.  Therefore 
 a naive dimensional analysis is expected to work.  Since the scaling dimensions 
 of the new parameters are 
\beqa
  &&[g_0] = -2, \ \ [g_1] = 6-d, \ \ [g_2]=[g_3]= 4-d, \nn\\
  &&[g_4]=2-d, \ \ [g_5] = -d 
\eeqa
respectively,  only $g_1$ is relevant near $d= 6$.   
This  suggests that the extra parameters $g_{2}, ..., g_{5}$ 
do not  play any important role for critical phenomena 
in the RFIM when $d$ is close to 6.  


\subsection{The Gaussian case}
In order to make our argument more precise, we need to show that 
 recursion relations for $(t, \{g_\mu\})$  do not contain positive
powers of $\Delta$ for {\em all} orders.

To this end, it is instructive to  consider the case where 
the probability distribution of impurities 
is Gaussian, i.e., $u_2, ..., u_5$ are ignored\cite{aim, g}. 
Suppose that  $u_1'$ is expressed as 
\beq
  u_1' 
  = L^{4 \theta -  3 d} \sum_{a_1=1}^{\infty} f_{a_1} \(\Delta, t \) u_1^{a_1}.  
\eeq
Here $f_{a_1} \(\Delta, t \)$ is obtained from the sum of all diagrams for $\rnd{v'_4}$ with 
$a_1$ $\phi^4$ vertices.   Since the number of $\times$ in a diagram gives the power of 
$\Delta$, $f_{a_1} \(\Delta, t \)$ is a polynomial of $\Delta$ with a finite degree. 
Letting $\gamma_{a_1}$ be that degree,  we can write 
\beq
  f_{a_1} \(\Delta, t \) = \sum_{n=0}^{\gamma_{a_1}} c_n \Delta^n, 
\eeq
where $c_n$ is a coefficient  that depends on $t$ and $a_1$.  Hence $u_1'$ is written as 
\beqa
  u_1' 
   &=& L^{4 \theta -  3d} \sum_{a_1=1}^{\infty} \sum_{n=0}^{\gamma_{a_1}}  
  c_{n} \Delta^n u_1^{a_1}
  \nn\\
  &=&L^{4 \theta - 3 d} 
  \sum_{a_1=1}^{\infty} \Delta^{\ga_{a_1}} u_1^{a_1} \sum_{n=0}^{\gamma_{a_1}}  
  c_n \Delta^{n-\ga_{a_1}}.  
\eeqa
Let us obtain $\gamma_{a_1}$.  Since the diagrams for $u_1'$ have a single connected 
component,  internal lines without $\times$ are needed at least $a_1-1$.  
Excluding four external lines from the total lines $4 a_1$,  
internal lines on which we can put $\times$ are at most 
\beq
  \frac{1}{2} \(4 a_1 -4 - 2 \( a_1 -1 \) \) = a_1-1. 
\label{numx}
\eeq
This is nothing but $\gamma_1$. 
Thus 
\beq
    u_1' = L^{4 \theta - 3 d} \Delta^{-1} \(\sum_{a_1=1}^{\infty}g_1^{a_1} 
	\sum_{n=0}^{a_1-1} c_n {g_0}^{\(a_1-1\)-n} \).
\label{u'1}
\eeq
Note that the power of $g_0$,  $a_1-1-n$, is 
non-negative in the above summation.   
Similar observations of  $\Delta' $ and $t'$ lead to the following form:
\beq
  \Delta'  = L^{2\theta - d} \Delta \( 1+ \sum_{a_1=1}^{\infty} g_1^{a_1}
  \sum_{n=0}^{a_1+1}  
  d_n g_0^{\(a_1+1\)-n}  \)
\label{delta'}
\eeq
and
\beq
  t' = L^{2\theta -d}\(t + 
	\sum_{a_1=1}^{\infty}g_1^{a_1} \sum_{n=0}^{a_1}  
  e_n {g_0}^{a_1-n} \)
\label{t'}
\eeq
where $\{d_n\}$ and $\{e_n\}$ are  $t$-dependent coefficients. 
From Eqs.(\ref{u'1}) and (\ref{delta'}), one finds that $g_1' = \Delta' u_1'$, 
$g_0' = (\Delta')^{-1}$ and $t'$ are expanded by $g_1$ and $g_0$ for all orders.

\subsection{General case}
Next, we extend the above argument to the case where the extra parameters
 $u_2, ..., u_5$ are included. 
Let ${\bf Z^+}$ be the set of non-negative integers.  Define 
\beqa
  I \equiv && \{a \, : \,  a = \(a_1, ..., a_5\), a_j \in {\bf Z}^+, 
  \nn\\
  && \forall j= 1, ..., 5, \, \sum_{j=1}^5 a_j \geq 1 \}.  
\eeqa
For $a \in I$,  we use the notation
\beq
  u^a \equiv \prod_{j=1}^5 u_j^{a_j}.  
\eeq
The recursion equation for $u_j$ can be written as 
\beq
  u'_j = L^{4\theta - 3d}  \sum_{a \in I} \Delta^{\ga^j_a} u^a \sum_{n=0}^{\gamma^j_a} c_n^j 
  \Delta^{n-\ga^j_a} .
\label{u'j}  
\eeq
Here we compute $\gamma^j_a$, the degree of $\Delta$ for $u^a$ in $u'_j$.  
The number of connected components in the vertex corresponding to $u_j$ is denoted by 
$\al_j$, namely, 
\beq
  \al_1 = 1, \ \ \al_2 = \al_3 = 2, \ \ \al_4 = 3, \ \ \al_5 = 4. 
\eeq

Here we consider a diagram proportional to $u^a$ in Eq. (\ref{u'j}). 
Before connecting the vertices in the diagram by internal lines, 
there are 
\beq
  \sum_{i=1}^5 \al_i a_i
\eeq
connected components.  
Since the diagram has $\al_j$ connected components after connecting the vertices,
 we need at least 
\beq
   \sum_{i=1}^5 \al_i a_i - \al_j
\eeq
no-$\times$ internal lines. Similar computation for Eq. (\ref{numx}) leads to  
\beqa
   \gamma^j_a &=&  
   \frac{1}{2} \(4 \sum_{l=1}^5 a_l -4 -2\(\sum_{i=1}^5 \al_i a_i - \al_j \)\) 
   \nn\\
   &=&  \be_j - \sum_{i=1}^5 \be_i a_i,
\label{gaja}
\eeqa
where we have defined $\be_j \equiv \al_j - 2$.  Explicitly, 
\beq
  \be_1 = -1, \ \ \be_2 = \be_3 = 0, \ \ \be_4 = 1, \ \ \be_5 = 2. 
\eeq
According to the definition of the expansion parameters, we find 
\beq
  g_j = \De^{-\be_j} u_j, \qquad j=1, ..., 5.
\label{gj}
\eeq
Applying the result of Eq. (\ref{gaja}) to Eq. (\ref{u'j}), we get 
\beq
  u'_j = L^{4\theta - 3d} \Delta^{\be_j} \sum_{a \in I} g^a 
  \sum_{n=0}^{\gamma^j_a} c_n^j  g_0^{\ga_a^j-n}.
\label{u'j2}
\eeq
Similarly, we find that $\Delta'$ and $t'$ are expressed by the following expansion: 
\beqa
   \Delta' &=&   L^{2\theta -d} \Delta\( 1+  \sum_{a \in I} g^{a}
  \sum_{n=0}^{\gamma_{a}}  
  \tilde{d}_n g_0^{\gamma_{a}-n}  \)\nn\\
  t' &=&  L^{2\theta -d} \(t +  \sum_{a \in I} g^{a}
  \sum_{n=0}^{\gamma_{a}}  
  \tilde{e}_n g_0^{\gamma_{a}-n}\)
\label{t&d}
\eeqa
with
\beq
  \gamma_a = - \sum_{i=1}^5 \be_i \al_i.  
\eeq
Using Eq. (\ref{u'j2}) and the above expansion for $\Delta'$, we conclude 
that $g'_j \equiv (\Delta')^{-\be j} u'_j$ is expanded by $\{g_\mu \}$.  
Further, it is obvious from Eq. (\ref{t&d}) that $g'_0$ and $t'$ also 
have perturbation series in terms of $\{g_\mu\}$. 
 
It should be noted that some physical quantities,  such as  free energy,  are proportional 
to the relevant parameter $\Delta$. Hence if we compute an exponent associated with 
those quantities, we have to know the singular behavior of $\Delta$ near the criticality.
As we can see in the first line of Eq. (\ref{t&d}), $\Delta$ is renormalized multiplicatively 
and its correction is expanded in terms of $\{g_\mu\}$.  Therefore, we can compute the
singular behavior within the framework of perturbation in terms of $\{g_\mu\}$.

\subsection{Observation for  dimensional reduction of the RFIM}
\label{observation}
 We have shown that the transformed parameters  $\{g'_\mu\}$ 
 can be expressed as  positive series in $\{g_\mu\}$.  Since the expansion 
 parameters other than $g_1$ are irrelevant near $d=6$, 
 we may ignore them.  
 
 As we have mentioned in Sec.  \ref{comp},  our recursion relations are 
 consistent with the replica method studied by 
 Br\'ezin and de Dominicis
 if we adopt the limiting procedure (A) explained in Sec.  \ref{comp}\cite{bd}. 
However,  it is concluded that the non-trivial fixed  point becomes 
unstable due to $u_2, ..., u_5$ in Ref.\cite{bd}.  
The discrepancy between this and our conclusion is  
resolved as follows:  in Ref.\cite{bd}, the recursion relations of  parameters 
$\tilde{g}_j \equiv \Delta u_j$ are computed. Since $[\tilde{g}_j] = 6 -d$ for all $j$, 
it is possible that some of $\tilde{g}_j$ becomes larger as the RGT is repeated.  
Nevertheless, it does not mean that  a diagram proportional to $\tilde{g}_j$ 
brings infrared divergence near the upper critical dimension. 

In fact,  if we write $\de t$ in Eq. (\ref{cor3}) by $\tilde{g}_j$ instead of $u_j$, 
 one can easily check that $\tilde{g}_j$s  for $j=2, ...,5$ are always 
 combined with $\Delta^{-1}$ or $\Delta^{-2}$.  Thus, even though $\tilde{g}_j$ 
 $(j=2,...,5)$ behave as $g'_j \sim L^{c_{j} \ep} g_j$ with $c_{j}>0$, the negative powers of 
$\Delta$ suppresses growth of terms proportional to $\tilde{g}_j$ in $\de t$.  
The discussion in the previous subsection shows that the association with the 
negative powers of $\Delta$ occurs for all orders.  Thus $\tilde{g}_j$s 
$(j=2, ..., 5)$ do not contribute to exponents.  

Although the extra parameters remain irrelevant near the upper critical dimension, 
it is plausible that those parameters become relevant when  $\epsilon$ exceeds some 
finite value $\epsilon_c$,  which can cause the breakdown of the dimensional 
reduction in $d=3$. The existence of such critical value is consistent with 
a high-temperature expansion by Houghton {\it et al}\cite{hks}.  
They concluded that dimensional reduction occurs in $d=5$ and 6 
while  the phase transition becomes first order in $d \leq 4$.   It is 
strongly suggested that $1< \epsilon_c < 2$. 
On the other hand, another high-temperature expansion performed by Gofman {\it et al} 
suggests that the breakdown of  dimensional reduction in $d \leq 5$\cite{gaahs}. 
It can be interpreted as $0< \epsilon_c < 1$.

As we explained above,  dimensional reduction can survive 
for sufficiently small $\epsilon$.  In this case,  the exponents 
$\nu$,  $\eta$ and $\bar{\eta}$ are calculated as\footnote{
  A derivation of these results are outlined in appendix \ref{scaling}. 
  }
\beq
  \nu = \frac{1}{2} + \frac{1}{12} \ep + {\rm O}(\ep^{2}), \ \ 
  \eta = \bar{\eta} = \frac{1}{54}\ep^{2} + {\rm O}(\ep^{3})
\eeq
However,  other parameters not considered here may 
bring the breakdown of 
dimensional reduction for all $d<6$.  
 It is conjectured by Feldman that the 
apparently higher-dimensional operators $\sum_{ab}(\phi^{a}-\phi^{b})^{2l}$ 
($l >1/\ep^{2}$) turn to relevant ones\cite{fe}. 
In the Harris-Lubensky method, similar operators 
$\sum_{ab}(\phi^{a}\phi^{b})^{l}$ are introduced,  
taking into account the following random average:
\beq
  \rnd{v_{l}\(k_{1}, ..., k_{l}\); v_{l}\(k'_{1}, ..., k'_{l}\)} = \bar{u}_{l} 
  \de\(k_{1} + \cdots + k'_{l} \). 
\eeq
The parameter $\bar{u}_{l}$ is, itself,  irrelevant.
However, following the argument in Sec. \ref{redef}, 
 one can  find   an expansion parameter 
 proportional to $\bar{u}_{l}$ is $\(\Delta\)^{l-2} \bar{u}_{l}$, which has 
the canonical dimension $(l-1)(4-d) \sim l (4-d)$ for large $l$.  
Feldman shows that  they acquire anomalous dimension with  
${\rm O}\(l^{2} \epsilon^{2} \)$ in the second-order perturbation.  
It means that  $\(\Delta\)^{l-2} \bar{u}_{l}$ can transmute a 
relevant parameter for sufficiently large $l$ satisfiying
 $l > 1/\ep^{2}$.  The conjecture 
should be checked by a non-perturbative means that  is beyond 
the scope of this paper.   Nevertheless it is consistent 
with our result in that the operator with $l=2$ 
(i.e., $\sum\(\phi^{a}\phi^{b}\)^{2}$) is irrelevant for sufficiently small 
$\epsilon$.

\section{Summary and discussion}
In this paper, we have studied the critical phenomena of the extended 
Ginzburg-Landau theory for the random-field Ising model. 
We have employed the renormalization group for the probability distribution of the impurities.  
Probability distribution is characterized by  non-trivial cumulants that  bring  
extra parameters, which are essentially identical to the new coupling constants in the 
replica Hamiltonian introduced in Refs.\cite{hks,bd}.   
In contrast to the replica method, our approach does not require any limiting procedures, and hence no artificial 
ambiguities arise.  Thus we can definitely determine 
the scaling dimensions of expansion parameters. 
We have found that  extra expansion parameters do not affect the critical phenomena in 
$d=6-\epsilon$ with sufficiently small $\epsilon$.   
On the other hand, those parameters could be an obstruction to the dimensional
 reduction at some finite $\epsilon$.  This result indicates that we cannot rule out 
  dimensional reduction near $d=6$ by including the extra coupling constants 
 in Eq. (\ref{rnd}).  It is consistent with the high-temperature expansion by
  Haughton {\it et al}\cite{hks}.  On the other hand, another high-temperature expansion by 
  Gofman {\it et al} is not consistent with Ref.\cite{hks}, which may indicate that the dimensional
   reduction does not occur in any $d<6$\cite{gaahs}.   
   It is important to resolve this discrepancy for clarifying mechanisms in  the phase transition 
   in the RFIM near the upper critical dimension. 
 
\begin{acknowledgments}
We wish to thank C. Itoi for valuable discussions.
\end{acknowledgments}

\appendix
\section{Computation of the recursion equations}
\label{recursion}
In this appendix, we show technical details of computing the correction terms Eq. (\ref{cor1})
in ${\rm O}\(u_i u_j \)$.  Eq. (\ref{cor3}) can be obtained in a similar manner. 

We do not obtain an explicit form of $v'_j$.  Instead, we derive all possible diagrams in the desired order 
that appear in the right-hand side of FIG. \ref{fig_pert}.  

\begin{figure}[h]
\begin{center}
\setlength{\unitlength}{1mm}
 \begin{picture}(70, 90)(0,0)
        \put(0,5){ 
		\includegraphics[width=70mm]{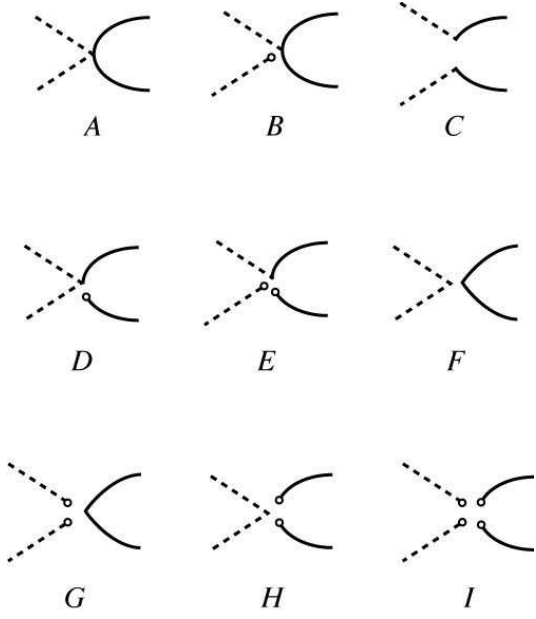}
		}
\end{picture}
\end{center}
\caption{nine vertices for $u'_j$ $(j=1, ..., 5)$}
\label{fig_nine}
\end{figure}

First, we consider diagrams without $\times$ on an internal line. 
As we have presented in FIG. \ref{fig_vertices}, the cumulants can be expanded by the five kinds of vertices.  
The lowest-order correction to $u'_j$ comes from the vertices with two external legs.  
There are nine such vertices as depicted in FIG. \ref{fig_nine}.  
Merging two of them, we obtain diagrams that contribute to $u'_j (j=1,...,5)$
up to the order ${\rm O}(u_i u_j)$.  
We write the number $j$ in 
Table \ref{table_oneloop} 
if the resulting diagram contributes to $u'_j$. 
\begin{table}[h]
\begin{tabular}{cccccccccc}
    &$A$&$B$&$C$&$D$&$E$&$F$&$G$&$H$&$I$\\
	$A$&1&2&1&1&2&3&4&3&4\\
	$B$&*&4&2&2&4&4&5&4&5\\
	$C$&*&*&3&2&4&3&4&4&5\\
	$D$&*&*&*&3&4&3&4\\
	$E$&*&*&*&*&5&4&5\\
	$F$&*&*&*&*&*\\
	$G$&*&*&*&*&*&*\\
	$H$&*&*&*&*&*&*&*\\
	$I$&*&*&*&*&*&*&*&*
\end{tabular}
\caption{Admissible one-loop diagrams made from the vertices in FIG. \ref{fig_nine}. If a diagram contributes to 
$u'_j$, the number $j$ is entered. }
\label{table_oneloop}
\end{table}
For instance, we can read from Table \ref{table_oneloop} that 
the diagram made of the vertices $A$ and $B$ contributes to $u'_2$.  
As we have mentioned in Sec.  \ref{diag}, any connected components  of 
a diagram in the right-hand side 
of FIG. \ref{fig_pert} must  contain at least one external line.  In Table \ref{table_oneloop}, 
a blank means that the corresponding diagram does not satisfy this condition. 

Now we calculate the numerical factor to each diagram, 
 as we have demonstrated 
in \ref{exam}. 
In contrast to the pure $\phi^4$ theory, 
 a one-loop diagram has multiple connected components 
in general.  There is the combinatoric factor associated with 
each connected component,  
which can be computed in the same way as with the pure theory.  
Here we denote the product of 
the combinatoric factor of each connected component as 
\beq
n_A.
\eeq 
Next, consider the case where a cumulant for $u_j$ has 
two (or more) identical potential  $v_i$. 
Suppose that $v_i'$ is expressed by the series $\sum_k v_i^{(k)}$.  If 
a diagram for $u'_j$ consists of $v_i^{(k)}$ and $v_i^{(k')}$ with $k \neq k'$, 
then 
\beqa
  u'_j &\propto&  \rnd{v'_i; \dots v'_i; \dots } \nn\\
         &=& \rnd{v_i^{(k)}; \dots v_i^{(k')}; \dots } + 
\rnd{v_i^{(k')}; \dots v_i^{(k)}; \dots } \nn\\
&&+ \cdots\nn\\
  &=&2 \rnd{v_i^{(k)}; \dots v_i^{(k')}; \dots } + \cdots.  
\eeqa
Such multiplicity is denoted as 
\beq
 n_R. 
\eeq
Finally, we have to take into account the number of ways of contraction in the random-potential 
average.  
Here it is denoted as 
\beq
  n_C.  
\eeq
Then, the factor with a diagram $n_F$ is computed as 
\beq
 n_F = n_A  n_R n_C.  
\eeq
 
In Tables \ref{table_u1}-\ref{table_u5}, we outline those factors.  
The first row denotes the power 
of $\Delta$ of the diagrams. The second row shows the name of the vertices in 
FIG. \ref{fig_nine} that are ingredients of the diagrams.  
The last row indicates the parameter 
dependence of the diagrams. 
\begin{table}[h]
  \begin{tabular}{|c|ccc|c|}
        \hline
        &\multicolumn{3}{c|}{$\Delta^0$}&\multicolumn{1}{c|}{$\Delta^1$}\\
        \hline
        &$AA$&$AC$&$AD$&$AA$\\
        \hline
	$n_A$&$-\frac{3}{2}$&6&6&$-3$\\
	$n_R$&1&1&1&1\nn\\
	$n_C$&1&1&1&1\nn\\
	$n_F$&$-\frac{3}{2}$ &6&6&$-3$\\
	\hline
	&$u_1^2$& $u_1^2$&$u_1u_3$&$u_1 u_2$\\
	\hline
  \end{tabular}
\caption{Diagrams for $u_1'$ and their numerical factors}
\label{table_u1}
\end{table}

%
\begin{table}[h]
  \begin{tabular}{|c|ccccc|cc|}
\hline
  &\multicolumn{5}{c|}{$\Delta^0$}&\multicolumn{2}{c|}{$\Delta^1$}\\
\hline
  &$AB$&$AE$&$BC$&$BD$&$CD$&$AC$&$AB$\\
\hline
  $n_A$&$-\frac{3}{2}$&$3$&$3$&$3$&$3$&$-3$&$-3$\\
  $n_R$&$1$&$1$&$1$&$1$&$1$&$1$&$1$\\
  $n_C$&$1$&$1$&$1$&$1$&$1$&$1$&$1$\\
  $n_F$&$-\frac{3}{2}$&$3$&$3$&$3$&$3$&$-3$&$-3$\\
\hline
 &$ u_1 u_2 $&$u_1 u_4$&$u_2 u_3$&$u_2^2$&$u_2 u_3$&$u_1u_3$&$u_1u_2$\\
\hline
  \end{tabular}
\caption{Diagrams for $u_2'$ and their numerical factors}
\label{table_u2}
\end{table}




\begin{table}[h]
  \begin{tabular}{|c|cccccc|cc|c|}
\hline
&\multicolumn{6}{c|}{$\Delta^0$}&\multicolumn{2}{c|}{$\Delta^1$}&
\multicolumn{1}{c|}{$\Delta^2$}\\
\hline
 &$AF$&$AH$&$CC$&$DD$&$CF$&$DF$&$AD$&$AF$&$AA$\\
\hline
$n_A$&$-\frac{1}{2}$&$\frac{1}{2}$&$1$&$1$&$1$&$1$&$-1$&$-1$&$\frac{1}{4}$\\
  $n_R$&$2$&$2$&$1$&$1$&$2$&$2$&$1$&$2$&$1$\\
  $n_C$&$1$&$1$&$2$&$1$&$1$&$1$&$2$&$1$&$2$\\
 $n_F$&$-1$&$1$&$2$&$1$&$2$&$2$&$-2$&$-2$&$\frac{1}{2}$\\
\hline
&$u_1 u_3$&$u_1 u_4$&$u_3^2$&$u_2^2$&$u_3^2$&$u_2u_3$&$u_1u_2$&$u_1u_3$&$u_1^2$\\
\hline
  \end{tabular}
\caption{Diagrams for $u_3'$ and their numerical factors}
\label{table_u3}
\end{table}




\begin{table}[h]
  \begin{tabular}{|c|ccccccccc|}
\hline
 &\multicolumn{9}{c|}{$\Delta^0$}\\
\hline
 &$AG$&$AI$&$BB$&$BE$&$BF$&$BH$&$CE$&$CG$&$CH$\\
\hline
 $n_A$&$-\frac{1}{2}$&$\frac{1}{2}$&$-\frac{1}{2}$&$2$&$-\frac{1}{2}$&$\frac{1}{2}$&$1$&$1$&$1$
\\
  $n_R$&$1$&$1$&$1$&$1$&$2$&$2$&$2$&$1$&$1$
\\
  $n_C$&$1$&$1$&$2$&$2$&$1$&$1$&$2$&$1$&$1$
\\
 $n_F$&$-\frac{1}{2}$&$\frac{1}{2}$&$-1$&$4$&$-1$&$1$&$4$&$1$&$1$
\\
\hline
&$u_1u_4$&$u_1u_5$&$u_2^2$&$u_2u_4$&$u_2u_3$&$u_2u_4$&$u_3u_4$&$
u_3u_4$&$u_3u_4$\\
\hline
\end{tabular}\\[3mm]

\begin{tabular}{|c|ccc|cccccc|}
\hline
&\multicolumn{3}{c|}{$\Delta^0$}&\multicolumn{6}{c|}{$\Delta^1$}\\
\hline
 &$DE$&$DG$&$EF$&$AE$&$BC$&$BD$&$CD$&$CC$&$CF$\\
\hline
$n_A$&$1$&$1$&$1$&$-1$&$-2$&$-1$&$-1$&$-1$&$-1$\\
$n_R$&$2$&$1$&$2$&$2$&$2$&$2$&$2$&$1$&$2$\\
$n_C$&$1$&$1$&$1$&$1$&$1$&$1$&$1$&$2$&$1$\\
$n_F$&$2$&$1$&$2$&$-2$&$-4$&$-2$&$-2$&$-2$&$-2$\\
\hline
&$u_2u_4$&$u_2u_4$&$u_3u_4$&$u_1u_4$&$u_2u_3$&$u_2^2$&$u_2u_3$&$u_3^2$&$
u_3^2$\\
\hline
\end{tabular}\\[3mm]

\begin{tabular}{|c|ccc|cc|}
\hline
&\multicolumn{3}{c|}{$\Delta^1$}&\multicolumn{2}{c|}{$\Delta^2$}\\
\hline
 &$AG$&$BB$&$BF$&$AC$&$AB$\\
\hline
$n_A$&$-1$&$-1$&$-1$&$-\frac{1}{2}$&$\frac{1}{4}$\\
$n_R$&$1$&$1$&$2$&$1$&$2$\\
$n_C$&$1$&$2$&$1$&$2$&$2$\\
$n_F$&$-1$&$-2$&$-2$&$1$&$1$\\
\hline
&$u_1u_4$&$u_2^2$&$u_2u_3$&$u_1u_3$&$u_1u_2$\\
\hline
\end{tabular}
\caption{Diagrams for $u_4'$ and their numerical factors}
\label{table_u4}
\end{table}
\begin{table}[h]
\begin{tabular}{|c|ccccc|cccc|}
\hline
&\multicolumn{5}{c|}{$\Delta^0$}&\multicolumn{4}{c|}{$\Delta^1$}\\
\hline
&$BG$&$BI$&$CI$&$E^2$&$EG$&$BE$&$CG$&$CE$&$BG$\\
\hline
$n_A$&$-\frac{1}{2}$&$\frac{1}{2}$&$1$&$1$&$1$&$-1$&$-2$&$-\frac{1}{2}$&$-1$\\
$n_R$&$4$&$4$&$6$&$6$&$4$&$12$&$12$&$12$&$4$\\
$n_C$&$3$&$3$&$1$&$2$&$3$&$2$&$1$&$2$&$3$\\
$n_F$&$-6$&$6$&$6$&$12$&$12$&$-24$&$-24$&$-12$&$-12$\\
\hline
&$u_2u_4$&$u_2u_5$&$u_3u_5$&$u_4^2$&$u_4^2$&$u_2u_4$&$
u_3u_4$&$u_3u_4$&$u_2u_4$\\
\hline
\end{tabular}\\[3mm]

\begin{tabular}{|c|ccc|}
\hline
&\multicolumn{3}{c|}{$\Delta^2$}\\
\hline
&$BC$&$CC$&$BB$\\
\hline
$n_A$&$\frac{1}{2}$&$1$&$\frac{1}{4}$\\
$n_R$&$12$&$1$&$6$\\
$n_C$&$2$&$6$&$4$\\
$n_F$&$12$&$6$&$6$\\
\hline
&$u_2u_3$&$u_3^2$&$u_2^2$\\
\hline
\end{tabular}
\caption{Diagrams for $u_5'$ and their numerical factors}
\label{table_u5}
\end{table}

It is worthwhile mentioning the relationship to the replica method\cite{bd}.  
From $S_{\rm int}^{\rm rep}$ defined in 
Eq. (\ref{srep}), 
\beqa
  \frac{\del^2 S^{\rm rep}_{\rm int}}{\del \phi_\al \del \phi_\be} 
  &=& 
  \frac{u^{\rm rep}_1}{2}\phi_{\al}^2 \de_{\al \be} + \frac{u^{\rm rep}_2}{2}
  \(\phi_\al^2 + \phi_\be^2\) 
  + u^{\rm rep}_3 \phi_\al \phi_\be
  \nn\\
  && + u^{\rm rep}_2 \de_{\al \be} \phi_\al \sum_{\mu}\phi_\mu + u^{\rm rep}_4 \(\phi_\al+\phi_\be\)
  \sum_{\mu}\phi_\mu 
  \nn\\
  &&+ \frac{u^{\rm rep}_3}{2} \de_{\al \be} \sum_{\mu}\phi_\mu^2+\frac{u^{\rm rep}_4}{2}\sum_\mu \phi_\mu^2 
  \nn\\
  &&+ \frac{u^{\rm rep}_4}{2} \de_{\al \be}\sum_{\mu \nu}
  \phi_\mu \phi_\nu + \frac{u^{\rm rep}_5}{2}\sum_{\mu \nu}\phi_\mu \phi_\nu
\eeqa
These terms exactly produce the vertices in FIG. \ref{fig_nine},  under the condition that 
each line  be connected in the case where the corresponding field has the common replica index. 
The one-loop diagrams identical with Table \ref{table_oneloop} are obtained using the replica propagator 
\beq
  G_{\al \be}(q) = \frac{\de_{\al \be}}{q^2+t} + \frac{\Delta}{(q^2+t)^2}. 
\eeq
We can check that the recursion equations identical with those in the main text 
are derived by this method,  if we take the limit  $n\rightarrow 0$ with $u_j$s 
fixed\cite{sak}.

\section{Critical exponents near the upper critical dimension}
\label{scaling}
In this appendix, we show an outline of computing critical exponents 
in $d=6-\epsilon$. 

Let  us introduce the connected and the disconnected two-point function 
$g_c(k; \mu)$ and $g_d(k; \mu)$ by the following formulae:
\beqa
  &&\rnd{\bra \phi\(k_1\); \phi\(k_2\)\ket} \equiv g_c\(k_1; \mu \) \de \(k_1+k_2\) 
  \nn\\
  &&\rnd{\bra \phi\(k_1\)\ket ; \bra\phi\(k_2\)\ket} \equiv g_d\(k_1; \mu \) 
  \de\(k_1+k_2\),  
\eeqa
where $\mu$ represents a point in the parameter space $\mu =\(t, g_0, ..., g_5\)$.  
Since $g_c\(k; \mu \)$ is computed from the sum 
of connected diagrams that appear in $\de t$,  it can be expanded by $\{g_\mu\}$.  
On the other hand, $g_d\(k; \mu \)$ is made of diagrams having two 
connected components appearing in $\de \Delta$.  Namely, $g_d(k; \mu)$ 
can be written as 
\beq
  g_d\(k; \mu \) = \Delta \tilde{g}_d\(k; \mu \), 
\eeq
where $\tilde{g}_d\(k; \mu \)$ is expressed as a 
perturbative series of $\{g_\mu\}$. 
Suppose that $\Delta$ transforms as 
\beq
  \Delta' \simeq L^\kappa \Delta. 
\eeq
The correlation functions  satisfy the following transformation law:
\beqa
  &&  g_c\(p; \mu \) = L^{2 \theta - d}g_c\(L p; \mu' \) \nn\\
  &&  \Delta \tilde{g}_d \(p; \mu \) = L^{2 \theta - d + \kappa}
  \Delta \tilde{g}_d\(L p; \mu' \).  
\eeqa
First,  we focus on the critical exponents $\eta$ and $\bar\eta$, which determine 
the small-momentum behavior of  correlation functions at the criticality:
\beqa
  g_c\(p; \mu \) &\simeq& \frac{1}{p^{2-\eta}} \delta\(p_1 + p_2\)
  \nn\\
  g_d\(k; \mu \) &\simeq&\frac{1}{p^{4-\bar\eta}} \de\(p_1+p_2\). 
\label{b-1}
\eeqa

 Suppose that we apply the RGT $n$ times at the criticality, where $n$ satisfies 
 \beq
  L^n p_1 = \Lambda  e_1, 
 \eeq
 with $e_1$ being some $d$ dimensional unit vector.  
The probability distribution 
$P$ approaches the fixed-point distribution, $P_*$,  
characterized by $\mu_*$.
We can evaluate $g_c(p; \mu)$ as 
 \beqa
   &&g_c(p_1; \mu) \nn\\
   &=& 
   L^{n (2 \theta-d)} g_c(L^n p_1; \mu^{(n)}) \nn\\
   &\simeq& \(\frac{\Lambda}{p_1} \)^{2 \theta -d} g_c(\Lambda e_1; \mu^*), 
  \eeqa
 where $\mu^{(n)}$ specifies the probability distribution of 
 impurities after having applied the RGT $n$ times. 
Comparing the definition (\ref{b-1}), we have 
\beq
  \theta = \frac{1}{2}\(2+d-\eta\). 
\label{b0.5}
\eeq
Applying the same argument to $g_d(p; \mu)$, we obtain 
\beq
  \theta = \frac{1}{2}\(4+d-\kappa -\bar\eta\) 
\label{b0}
\eeq
and
\beq
  \kappa = 2 + \eta -\bar\eta. 
\label{kappa}
\eeq

The exponents $\gamma$ and $\bar{\gamma}$ are  computed,  respectively,  
from 
$g_c(0; \mu)$ and $g_d(0;\mu)$ in the disordered phase. In this case, 
the RGT is repeated $n$ times, 
where $L^n$ is equal to the correlation length $\xi$.  Then 
\beqa
  \chi &=& g_c(0; \mu) \nn\\
  &\simeq& g_c(0;  \mu^{(n)}_i) \xi^{2\theta -d}.  
\eeqa
Assuming that 
\beq
  \xi \simeq (t - t_c)^{-\nu}, 
\eeq
we get, with the help of Eq. (\ref{b0.5}), 
\beq
  \chi \simeq \(t - t_c\)^{-\gamma}, 
\eeq
where 
\beq
  \gamma = (2-\eta) \nu. 
\eeq
Similar arguments for $g_d(0; \mu)$ lead to 
\beq
  \bar\chi = g_d(0; \mu ) \simeq (t - t_c)^{-\bar\gamma}
\eeq
with 
\beq
  \bar\gamma = (4 -\bar{\eta}) \nu. 
\eeq

Next we consider the exponent $\alpha$.  The singular part of  
free energy $F(\mu)$ transforms as 
\beq
  F(\mu ) = F(\mu'),   
\eeq
hence its density $f$ transforms as 
\beq
  f(\mu') \equiv \frac{F(\mu' )}{V'} = \frac{F(\mu)}{L^{-d} V} = L^d f(\mu), 
\eeq
where $f(\mu)$ has the form of 
\beq
  f(\mu) = \Delta \tilde{f}(\mu) 
\eeq
with $\tilde{f}(\mu)$ having the perturbative series of $\{g_\mu\}$.  
Thus 
\beq
 f(\mu) \simeq L^{(-d + \kappa) n} f(\mu^{(n)}) \simeq \xi^{-d + \kappa}, 
\label{fsing}
\eeq
which means that 
\beq
  2 -\al = \(d-2-\eta +\bar\eta\).  
\eeq
Eq. (\ref{fsing}) shows that $\ka$ is identical with the exponent of  
the singular part of the free energy in a correlation volume $\xi^d$, which is 
often denoted by $\theta$\cite{n,g,gaahs}.

In $d= 6-\epsilon$, we can perform the perturbation explicitly\cite{im,aim,g} . 
We begin with the equality 
\beqa
   \(k_1^2 + t' \) \de \(k_1 + k_2 \) &=& \rndp{v_2\(k_1, k_2\)}\nn\\
   &=& \rnd{v'_2\(k_1, k_2\)}. 
\label{b1}
\eeqa
Employing Eq. (\ref{cor2}) and denoting 
\beq
  \rnd{\de v_2 \(p_1, p_2\)} = \de \Gamma_2\(p_1; \mu \) \de \(p_1 + p_2\), 
\eeq
we get 
\beqa
  &&\rnd{v'_2\(k_1, k_2\)} = \nn\\
  &&L^{2\theta -d} 
   \(k_1^2L^{-2} + t + \de \Gamma_2 \(p_1;\mu \) \) \de \(k_1 + k_2\). 
\label{b2}
\eeqa
Comparing the coefficient of $k_1^2$ in Eqs.(\ref{b1}) and (\ref{b2}), we get 
\beq
  L^{\eta} = 1 + \left. \frac{\del }{\del p_1^2} \right|_{p_1=0} \de \Gamma_2 \(p_1^2; \mu_*\), 
\label{b3} 
\eeq
which determines the exponent $\eta$.  On the other hand, $\bar\eta$ is computed by a 
correction to $\Delta$.  Define $\de \Gamma_{\Delta}(p;\mu)$ 
 by 
\beqa
  &&\rnd{v'_1\(k_1\) ; v'_1\(k_2\)} = 
  \nn\\
  &&L^{2\(\theta-d\)}\Delta
  \( 1+ \de\Gamma_{\Delta}\(p_1;\mu\)\) \de\(p_1 + p_2\). 
\eeqa
We can readily derive  
\beq
\Delta' = L^{2\theta -d } 
 \Delta\( 1 + \de \Gamma_{\Delta} \(0; \mu_*\) \).  
\eeq
Repeating similar calculations for deriving Eq. (\ref{b3}), we obtain 
\beq
  L^{2 \eta -\bar{\eta}} =  1 + \de \Gamma_{\Delta} \(0; \mu_*\). 
\label{b4} 
\eeq

When $d=6-\epsilon$, we can evaluate $\de \Gamma_2$ and $\de \Gamma_{\Delta}$  by 
the $\epsilon$ expansion.  
Here we can ignore the irrelevant parameters $g_2, ..., g_5$.  
In this case, we have 
\beq
  \left. \frac{\del }{\del p_1^2} \right|_{p_1=0} \de \Gamma_2 \(p_1^2; \tilde\mu\) 
  =  \de \Gamma_{\Delta} \(0; \tilde\mu\), 
\eeq
where $\tilde\mu$ is a point in the parameter space satisfying $g_2=\cdots =g_5 = 0$.  The above equation 
indicates that 
\beq
  \eta = \bar{\eta}
\label{eeta}
\eeq
with Eqs. (\ref{b3}) and (\ref{b4}).  This equality shows that 
\beq
  \kappa = 2
\eeq
from Eq. (\ref{kappa}).  It derives 
\beq
  g'_1 \simeq L^{\kappa + 4-d - 2\eta} \(g_1 + \de g_1 \) 
  = L^{6-d - 2\eta} \(g_1 + \de g_1 \),  
\eeq
where  $\de g_1$ is shown to be identical with that of the $4 -\epsilon$ dimensional 
pure $\phi^4$ theory,  with the coupling constant $g_1$\cite{aim}.  Further,   the correction 
term $\delta t$ is also equal to that of the pure theory in $4-\epsilon$ dimensions.  In this 
way,  we can rederive the result of  dimensional reduction near the upper critical dimension. 

\end{document}